\documentclass[twocolumn,showpacs,amsmath,amssymb]{revtex4}

\usepackage{epsfig,psfrag}
\usepackage{dcolumn}% Align table columns on decimal point
\usepackage{bm}% bold math
\usepackage{graphicx}
\usepackage{color}
\newcommand{\beq}{\begin{equation}}
\newcommand{\eeq}{\end{equation}}
\newcommand{\beqr}{\begin{eqnarray}}
\newcommand{\eeqr}{\end{eqnarray}}
\newcommand{\e}{{\epsilon}}

\newcommand{\psib}{{\bar{\psi}}}

\def\bS{{\mathbf S}}

\def\tJ{{\tilde{J}}}

\def\ua{\uparrow}
\def\da{\downarrow}
\def\eqa{\begin{eqnarray}}
\def\eea{\end{eqnarray}}

\def\cD{{\cal D}}

\def\cT{{\cal T}}

\def\cs{{\cal S}}
\def\ve{{\varepsilon}}
\def\a{{\alpha}}
\def\b{{\beta}}

\def\lol{{L\over\xi}}
\def\sgn{{\mathrm sgn}}

\def\eurlet{Europhys. Lett.}

\begin{document}

\title{A Universal Interacting Crossover Regime in Two-Dimensional Quantum Dots}
\author{Ganpathy Murthy} 
\affiliation{Department of Physics and Astronomy,
University of Kentucky, Lexington KY 40506-0055} \date{\today}
\begin{abstract}
Interacting electrons in quantum dots with large Thouless number $g$
in the three classical random matrix symmetry classes are
well-understood. When a specific type of spin-orbit coupling known to
be dominant in two dimensional semiconductor quantum dots is
introduced, we show that a new interacting quantum critical crossover
energy scale emerges and low-energy quasiparticles generically have a decay width
proportional to their energy. The low-energy physics of this system is
an example of a universal interacting crossover regime.

\end{abstract}
\vskip 1cm 
\maketitle

The statistics of the single-particle states of mesoscopic systems
with disorder or chaotic boundary
scattering\cite{mesoscopics-review,qd-reviews} are controlled by
Random Matrix Theory (RMT)\cite{rmt}, as long as the states are
separated by less than the Thouless energy $E_T$ (related to the
ergodicization time for a particle $\tau_{erg}$ by the Uncertainty
Principle $E_T=\hbar/\tau_{erg}$). For mean single-particle level
spacing $\delta$, the Thouless number is $g=E_T/\delta$.

Since disorder breaks all the spatial symmetries, only time-reversal
$\cT$ and possibly Kramers degeneracy remain. There are three
classical symmetry classes\cite{rmt}, the gaussian orthogonal ensemble
or GOE ($\cT$ intact, no spin-orbit coupling), the unitary or GUE
($\cT$ broken), and the symplectic or GSE ($\cT$ intact, with
spin-orbit coupling). More recently, other classes have been
identified for disordered superconductors\cite{zirnbauer} and quantum
dots constructed from two-dimensional semiconductor heterostructures
with spin-orbit coupling\cite{aleiner-falko}. We will focus on a
symmetry class in the latter case (which we call the Aleiner-Falko
(AF) class\cite{aleiner-falko}) in which, after a canonical
transformation, the single-particle $S_z$ is conserved, while
$\bS^2$ is not\cite{aleiner-falko}.

The idea of a crossover between two symmetry classes will play a
central role in this paper.  Consider a system in which the
Hamiltonian is crossing over from the GOE to the GUE\cite{rmt},
acheived, e.g., by turning on the orbital effects of a magnetic
field. For Thouless number $g\gg1$ of the original GOE, the $g\times g$ crossover
Hamiltonian is
\beq
H_{X}(\a)= H_{GOE}+{\a\over \sqrt{g}} H_{GUE}
\label{hcross}\eeq
where $\a$ is the crossover parameter. Properties of
eigenvector\cite{cross-single,adam-x} correlations have been computed
in the crossover. For $g\gg 2\a^2=g_X\gg1$, the following
ensemble-averaged correlations hold for the eigenstates
$\psi_{\mu}(i)$, where $\mu\ne\nu$ label the states and $i,j,k,l$ the
original orthogonal labels:
\beqr
\langle\psi^*_{\mu}(i)\psi_{\nu}(j)\rangle =&{1\over g} \delta_{\mu\nu}\delta_{ij}\nonumber \\
\langle\psi^*_{\mu}(i)\psi^*_{\nu}(j)\psi_{\mu}(k)\psi_{\nu}(l)\rangle=&{\delta_{ik}\delta_{jl}\over g^2}+{\delta_{ij}\delta_{kl}\over g^2} {{E_{X}}\delta/\pi\over {E_{X}}^2+ (\e_{\mu}-\e_{\nu})^2}
\label{cross-correlations}\eeqr
The last term on the second line shows the extra correlations induced
in the crossover\cite{adam-x}. The crossover scale
${E_{X}}=g_X\delta/\pi$ represents a window within which GUE
correlations have spread, while GOE correlations remain at high
energies.

The crossover from the spin-rotation invariant GOE to the AF class is
a GOE $\to$ GUE crossover, where the ``magnetic flux'' has opposite
signs for opposite eigenvalues of $S_z$\cite{aleiner-falko}. If the
linear size of the system is $L$ and the spin-orbit scattering length
is $\xi\gg L$, this new AF symmetry class manifests itself below
$E_X\simeq (\lol)^4 E_T$. The crossover to the fully symplectic GSE
occurs\cite{aleiner-falko} at the parametrically smaller energy scale
of $(\lol)^6 E_T$, set here to 0.

Turning from single-particle physics to interactions, for small to
moderate $r_s$ the ``Universal Hamiltonian''\cite{H_U,univ-ham} is
known to contain all the relevant couplings\cite{qd-us1} at low
energies in the renormalization group sense\cite{rg-shankar}.
\beq
H_U=\sum\limits_{\a,s}\e_{\a}c^{\dagger}_{\a,s}c_{\a,s}+{U_0\over 2}{\hat N}^2 -J\bS^2+\lambda T^{\dagger} T
\label{univ-ham}\eeq 
Here ${\hat N}$ is the total particle number, $\bS$ is the total spin,
and $T=\sum c_{\b,\da}c_{\b,\ua}$. $H_U$ has a charging energy $U$, an
exchange energy $J$ and a superconducting coupling $\lambda$. This
last term is absent in the GUE, while the exchange term disappears in
the GSE. In the large-$g$ limit only interaction terms which are
invariant under the symmetries of the one-body Hamiltonian appear in
$H_U$\cite{H_U,univ-ham}. At larger $r_s$, the system enters a quantum
critical regime\cite{critical-fan} connected with the impending
Pomranchuk transition\cite{qd-us2,qd-long}.

In the GOE, ignoring the Cooper coupling (two-dimensional
semiconductor quantum dots do not superconduct) and tuning $J$ one
sees the mesoscopic Stoner effect\cite{H_U,univ-ham}. Take for
illustration an evenly spaced set of levels with level spacing
$\delta$. Since both $\bS^2$ and $S_z$ are conserved, we can focus on
the ground state with $S_z=S$, and find its energy for an even number
of particles to be $E_{gs}(S)=S^2\delta-JS(S+1)$. Defining
$\tJ=J/\delta$ and minimizing with respect to $S$ leads to steps from
$S=0$ to $S=1$ at $\tJ={1\over2}$, from $S=1$ to $S=2$ at
$\tJ={3\over4}$ etc. Including the mesoscopic (sample-to-sample)
fluctuations of the energies and the matrix elements leads to
probability distributions for spin
$S$\cite{H_U,univ-ham,alhassid-malhotra}. Note that the $S_z=S$ ground
 state is electronically uncorrelated (a single Slater
determinant, not a superposition).

What happens when spin-orbit coupling of the AF class is introduced
into a slightly generalized form of $H_U$\cite{alhassid-rupp}?
\beq
H=\sum \e_{\mu s}c^{\dagger}_{\mu s} c_{\mu s} -J_z S_z^2 -J(S_x^2+S_y^2)
\label{startH}\eeq
Here the basis $\mu$ labels the eigenbasis of the single-particle
AF crossover hamiltonian\cite{aleiner-falko}, and $s$ is the eigenvalue
of $S_z={\hbar\over2}\sum
c^{\dagger}_{\mu\uparrow}c_{\mu\uparrow}-c^{\dagger}_{\mu\downarrow}c_{\mu\downarrow}$.
$S_{x,y}$ are expressed\cite{alhassid-rupp} in terms of the
combinations $S_{\pm}=S_x\pm iS_y$ as
\beq
S_+=\sum\limits_{\a\b}M_{\a\b}c^{\dagger}_{\a\ua}c_{\b\da}
\eeq
with $S_-=(S_+)^\dagger$.  The matrix $M_{\a\b}$ depends on the
particular realization, and has the ensemble average\cite{alhassid-rupp}
\beq
\langle|M_{\a\b}|^2\rangle={{E_{X}}\delta/\pi\over {E_{X}}^2+(\e_{\a\ua}-\e_{\b\da})^2}
\label{Mensave}\eeq
At strong spin-orbit coupling\cite{H_U,univ-ham,alhassid-rupp}, noting
that $-J(S_x^2+S_y^2)$ is irrelevant in the RG
sense\cite{qd-us1,qd-us2,cross-us}, we end up with the Hamiltonian of
Eq. (\ref{startH}) with $J=0$. The ground and excited states of this
Hamiltonian have definite $S_z$ and are electronically
uncorrelated.
\begin{figure}
\epsfxsize=2.9in\epsfysize=2.8in \hskip
0.3in\epsfbox{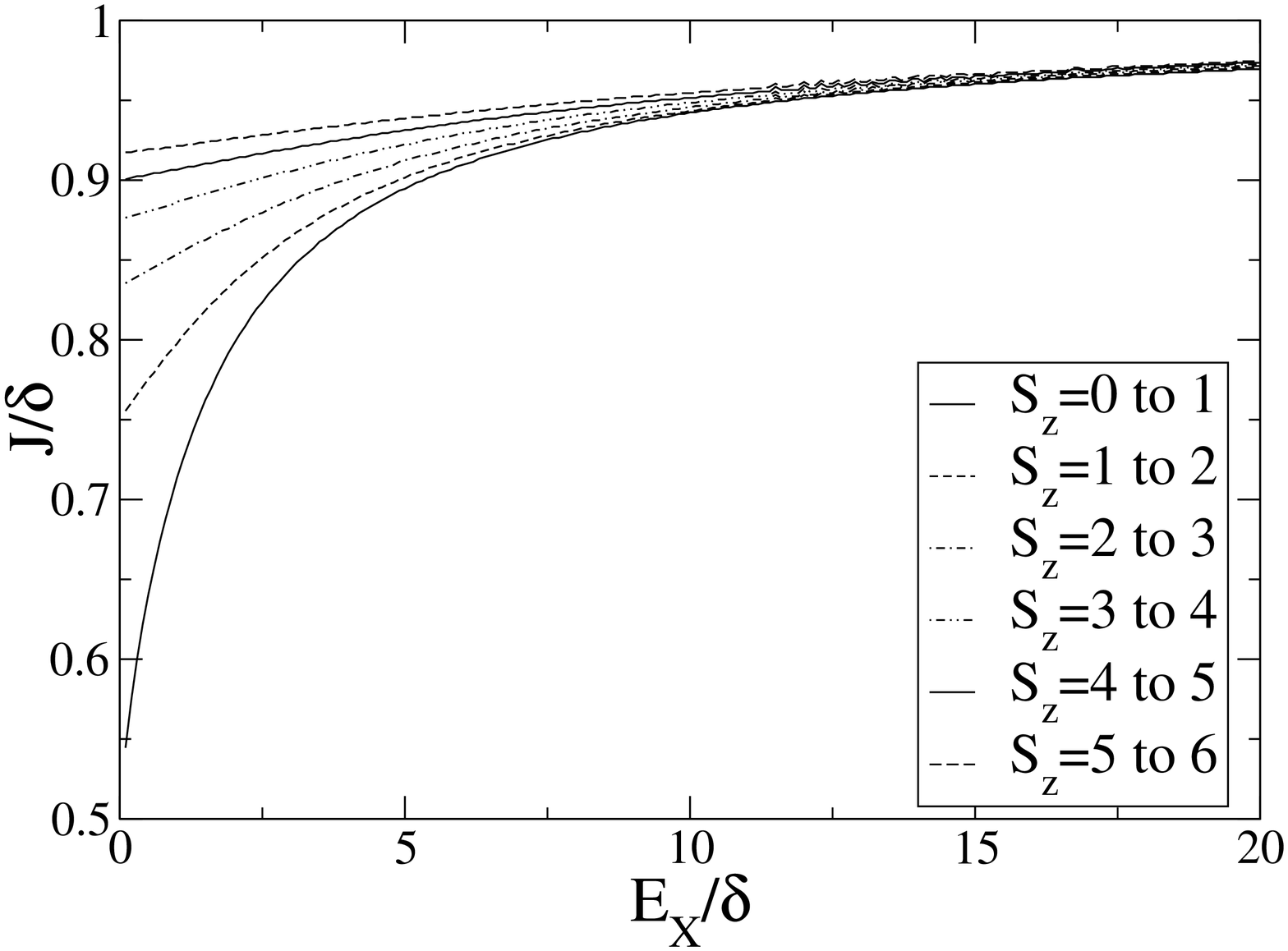} \vskip 0.1in \caption{Phase diagram in the
  $E_X-\tJ$ plane showing the lines of ground state $S_z$ 
  transitions. Note the smooth approach to the spin-rotation-invariant
  result as $E_X\to0$. } \label{fig1}
\end{figure}

We are now ready to state our central results. In the crossover to the
AF class, a quantum critical regime\cite{critical-fan} emerges at a
many-body quantum critical crossover scale $E_{QCX}=(1-\tJ)E_X$ (a
result the author obtained previously\cite{cross-us} in the restricted
case $J_z=0$). {\it In contrast to the limits of zero and strong
spin-orbit coupling, in the crossover the ground and low-lying states
are electronically strongly correlated. Transverse spin fluctuations
have a nonzero density at low energies  and
fermionic quasiparticles become very broad at low energies (as long as ground state $S_z\neq 0$)}. Finally,
the mesoscopic Stoner effect is smoothly pushed to higher $J$ as
spin-orbit coupling increases.

We will set $J_z=J$ in Eq. (\ref{startH}) henceforth, since that is
the correct starting point for a Hamiltonian deep in the Thouless
band\cite{qd-us1,qd-us2,qd-long}.  We decompose both the interaction terms by
introducing the Hubbard-Stratanovich fields $h(t)$, $q(t)=X(t)+iY(t)$,
and $q^*(t)$ to get the $T=0$ imaginary time action
\beqr
&A=\int\limits_{-\infty}^{\infty} dt\sum\limits_{\a,ss'}\psib_{\a s}[(\partial_t+\e_{\a})\delta_{ss'}-{h\over2}(\sigma_z)_{ss'}]\psi_{\a s'}\nonumber\\
&+{h^2+qq^*\over4J}-{q(t)\over2}\sum\limits_{\a\b}M_{\a\b}\psib_{\a\ua}\psi_{\b\da}-{q^*(t)\over2}\sum\limits_{\a\b}M_{\a\b}^*\psib_{\b\da}\psi_{\a\ua}\label{Soriginal}
\eeqr
$X(t),Y(t)$ are fluctuating fields and are integrated out, but $h(t)$
acquires an expectation value (also
called $h$). For our picket fence spectrum $\e_n=\delta(n-{1\over2})$
with chemical potential at $0$, at $h=0$ all states $n\le0$ are
occupied while all states $n\ge1$ are empty. When $h$ lies between
$h_n=\delta(2n-1)$ and $h_{n+1}=\delta(2n+1)$, at $T=0$ the
single-particle states between $-n+1$ and $n$ singly occupied, and
$S_z=n\equiv \cs$. We will find saddle points for $h$ in each of the
intervals $h_n< h< h_{n+1}$, thereby obtaining the ground state energy
as a function of $\cs=n$, and look for the lowest one.

First integrate out the fermion fields and obtain a quadratic
effective action for $h$, $X$, and $Y$.
\beqr
A_{eff}=&\int\limits_{-\infty}^{\infty} {d\omega\over2\pi} {|h(\omega)|^2\over4J}+{|X(\omega)|^2+Y(\omega)|^2\over4J}(1-J\chi_R(i\omega))\nonumber\\
&+{X(\omega)Y(-\omega)-Y(\omega)X(-\omega)\over4J}J\chi_I(i\omega)
\label{seff1}\eeqr
The cross terms are a consequence of $[S_x,S_y]=iS_z$, and $\chi_R$
and $\chi_I$ are the real and imaginary parts of $\chi$, the fermionic
transverse spin-susceptibility.
\beq
\chi(i\omega)=\sum\limits_{mn}|M_{mn}|^2{N_F(\e_{m\ua})-N_F(\e_{n\da})\over i\omega+\e_{n\da}-\e_{m\ua}}
\eeq
where $N_F$ is the Fermi occupation. To make further progress, we
assume that $E_X\gg\delta$, which allows us to convert the sums over
states into integrals, and also replace the sample-specific value of
$|M_{mn}|^2$ by its ensemble average. Such self-averaging occurs
naturally in the large-$N$ limit\cite{qd-us2,qd-long}. The dominant
contribution to $\chi$ is 
\beq
\chi(i\omega)={1\over\delta}{E_X-iE_{\cs}\ \sgn(\omega)\over |\omega|+E_X-ih\ \sgn(\omega)}
\eeq
where $E_\cs=2\cs\delta$ and $\sgn(\omega)$ is the sign of $\omega$.
Next, the integration over $X$ and $Y$ results in a fluctuation 
contribution to the effective action
\beq
A_{fluc}=\int\limits_{-\infty}^{\infty} {d\omega\over4\pi} \log\bigg({(|\omega|+E_{QCX})^2+(h-\tJ E_{\cs})^2\over(|\omega|+E_X)^2+h^2} \bigg)
\eeq
where the argument of the logarithm is the determinant of the matrix
of the quadratic form of $X$ and $Y$. $A_{fluc}$ is
logarithmically divergent  due to a limitation
of the ensemble averages Eqs. (\ref{cross-correlations},\ref{Mensave})
for large energy separations $E_X> E_T$.  Cutting it off, discarding
terms of the form $\log(E_T/E_X)$ which are independent of
$\cs$, defining $\Delta h=h-\tJ
E_{\cs}$, and adding the one-body energy we find
\begin{widetext}
\beq
A_{eff}(\cs,h)=\cs^2\delta+{h^2\over4J}-h\cs+{1\over\pi}
\bigg(\Delta h\tan^{-1}{\Delta h\over E_{QCX}}-h\tan^{-1}{h\over E_X} + {E_X\over2}\log(1+{h^2\over E_X^2})-{E_{QCX}\over2}\log(1+{(\Delta h)^2\over E_{QCX}^2})\bigg)
\label{seff2}\eeq
\end{widetext}
Eq. (\ref{seff2}) is one of the central results of this paper. Note
that the term in the brackets is or order $1/\cs$ compared to the
first three terms ($h$ will turn out to be order $\cs$).  From it we
find that the saddle point value of $h_0$ satisfies
\beq
h_0=\tJ E_S+{2J\over\pi}\big(\tan^{-1}{\Delta h_0\over E_{QCX}}-\tan^{-1}{h_0\over E_X}\big)
\label{hsp}\eeq
We must also ensure that $h_n<h_0<h_{n+1}$ for $\cs=n$. Using $h=h_0$
in Eq. (\ref{seff2}) gives the ground state energy for that
$\cs$. Even though this result has been derived for $E_X\gg\delta$ and $\cs\gg1$,
note that one recovers the correct spin-rotation-invariant result for all $\cs$ on
taking the $E_X\to0$ limit first in Eq. (\ref{seff2}) and solving it
to obtain the ground state energy.  Fig. \ref{fig1} shows regions in
the $E_X,\ \tJ$ plane with different ground state $S_z=\cs$. Note the
smooth approach to the spin-rotation invariant results as $E_X\to0$.

Now consider the low-lying excitations, which are bosonic spin
excitations and fermionic quasiparticles. The transverse spin correlator 
\beq
\cD(t)=-\langle T_tS_+(t)S_-(0)\rangle=-{1\over4J^2}\langle T_t q^*(t) q(0)\rangle+{1\over J}
\eeq
as a function of Matsubara frequency (at $T=0$) is
\beq
\cD(i\omega)=-{1\over\delta}{E_X+iE_S\ \sgn(\omega)\over|\omega|+E_{QCX}+i\Delta h\ \sgn(\omega)}
\eeq
The last term in the denominator is never more than a fraction of
$\delta$ and will be ignored below. Going over to the retarded
commutator with the standard replacement $i\omega\longrightarrow
\omega+i0^+$, we find the spectral function for transverse spin excitations
\beq
B(\omega)=-2Im(D_{ret}(\omega))={2\over\delta}{\omega E_X+E_S
E_{QCX}\over\omega^2+E_{QCX}^2}
\label{B}\eeq
For $\cs=0$ this reproduces the scaling function computed
earlier\cite{cross-us}. More importantly, for $\cs\neq0$ there is a
nonzero density of spin excitations even as ${\omega\over
E_{QCX}}\ll1$ (but $\omega\gg\delta$). These excitations are related
to the $2S+1$-degenerate ground states of the spin-rotation-invariant
system (since $\bS^2$ is not conserved here, the ground state contains
a superposition of many values of $\bS^2$). These low-energy
excitations make the system strongly correlated in the electronic
sense, that is, the ground and low-lying states are no longer
described even approximately by single Slater determinants.

The decay of an electron of energy $\ve$ to leading
order occurs via the emission of a single spin excitation. Using
the Fermi Golden Rule, we get the decay rate 
\beq
\Gamma(\ve)\simeq 2\pi J^2\int\limits_{0}^{\ve} d\ve'|M(\ve-\ve')|^2 B(\ve-\ve')\rho(\ve')
\eeq
where $\rho(\ve')$ is the electronic density of states at energy
$\ve'$ and $M$ is the matrix of Eq. (\ref{Mensave}).  Using
$\rho(\ve')=1/\delta$, and Eqs. (\ref{Mensave},\ref{B}) we obtain
\begin{widetext}
\beq
\Gamma(\ve)\simeq {J^2E_X\over\delta\pi(E_X^2-E_{QCX}^2)}\bigg({E_X\over2}\big[\log(1+{\ve^2\over E_{QCX}^2})-\log(1+{\ve^2\over E_X^2})\big]+4E_SE_{QCX}\big[{\tan^{-1}{\ve\over E_{QCX}}\over E_{QCX}}-{\tan^{-1}{\ve\over E_{X}}\over E_{X}}\big]\bigg)
\eeq
\end{widetext}
This is the other central result of this paper.  At low energies
$\ve\ll E_{QCX}$, $\rho$, $|M|^2$, and $B$ are constant, leading to a
decay rate which goes as $\Gamma(\ve)\simeq {16J^2\cs\over
E_XE_{QCX}}\ve$, {\it which can exceed $\ve$ for $\tJ\to1$, leading to
ill-defined quasiparticles}.  At high energies $\ve\gg E_X$ (but
$\ve\ll E_T$), the decay rate due to spin excitations goes to a
constant. There will be an additional Fermi liquid broadening\cite{fock-deloc}, $\Gamma\simeq \ve^2/E_T$ due to
neglected interactions.  Thus we have an unusual situation in which
the quasiparticles are better defined at high energies than at low
energies, because the high-energy physics is controlled by the weakly
interacting spin-rotation-invariant $H_U$\cite{H_U,univ-ham}.

In summary, we have established that a crossover from the
spin-rotation-invariant GOE to the AF class\cite{aleiner-falko} in
two-dimensional semiconductor quantum dots makes the ground and
low-lying states of the system strongly correlated in the electronic
sense, with a nonzero density of states for transverse spin
excitations at very low energies, and a decay rate
$\Gamma(\ve)\simeq\ve$ for low-energy fermionic quasiparticles (for
ground state $S_z\ne0$). The results are universal since all energies
are $\ll E_T$, and apply to disordered as well as ballistic/chaotic
dots.

Random matrix crossovers thus offer us access to {\it universal interacting
crossover regimes}\cite{cross-us}. Such regimes are dominated by
many-body correlations and are distinct from single-particle RMT
ensembles\cite{rmt,zirnbauer,aleiner-falko}, and offer the possibility of tuning  the
many-body quantum crossover scale $E_{QCX}$ by varying the single-particle $E_X$.

Another example of such a universal interacting crossover regime is a
superconducting dot vertically tunnel-coupled to a normal dot with an
orbital flux\cite{two-dots}. The crossover scale is varied by
varying the tunneling strength. In our problem one can
imagine a pair of vertically coupled quantum dots with one of them
fabricated of a different material having a much larger  spin-orbit
coupling than the other. $E_X$ (and thus $E_{QCX}$) would be tuned by varying the tunneling strength.

Let us turn to some caveats. We have focused on an even number of
electrons in the dot but expect similar results for an odd number. Our
large-$N$ approach is valid for $E_X\gg\delta$ and $\cs\gg1$. However, even in this
regime, in a tiny region of width $\Delta\tJ\simeq\delta^2/(E_X\cs)$
around the transition between $S_z$ steps, the value of $h_0$ is such
that the fermionic susceptibility exceeds $1/J$, and a small
saddle-point value for $|q|$ will be generated, leading to a more
complicated effective action. Apart from potentially missing effects
in this region, the large-$N$ approach seems to capture most of the
physics.  We have assumed a spectrum with equal level spacing, and the
ensemble averaged form for the matrix elements $M_{\a\b}$. In real
dots in the crossover, both
fluctuate\cite{rmt,cross-single,adam-x}. One thus expects mesoscopic
(sample-to-sample) fluctuations in the ground state value of
$S_z=\cs$\cite{H_U,univ-ham,alhassid-malhotra,alhassid-rupp} leading
to a probability distribution for $\cs$ given $\tJ$ and $E_X$. Our 
methods  can be used to complement exact diagonalizations\cite{hakan-yoram} in this case.

Many open questions remain, such as how to characterize the state at
and near the transition between $S_z$ steps, how the state responds to
an in-plane $B$ field, whether the strong electronic correlation has
any signatures in the zero-bias conductance, and whether one can
classify universal interacting crossover regimes into universality
classes. The author hopes to explore these and other questions in
future work.

It is a pleasure to thank R. Shankar and and especially Yoram Alhassid
for comments on the manusript.

\end{document}